\documentclass[iop,tighten]{emulateapj}
\usepackage{amsmath}
\usepackage{mathpazo}
\usepackage{cancel}
\newcommand{\Bx}[0]{\mathbf{x}}
\newcommand{\Bv}[0]{\mathbf{v}}
\newcommand{\By}[0]{\mathbf{y}}
\newcommand{\T}[0]{{\text{T}}}
\DeclareMathOperator{\Proj}{Proj}
\newcommand{\Abs}[1]{{\left\lvert{#1}\right\rvert}}
\newcommand{\inv}[1]{\frac{1}{#1}}
\newcommand{\Norm}[1]{\left\lVert{#1}\right\rVert}

\newcommand{\chapter}[1]{\title{#1}}
\newcommand{\blogpage}[1]{}
\newcommand{\revisionInfo}[1]{}

\newcommand{\beginArtWithToc}[0]{
\begin{document}}
\newcommand{\beginArtNoToc}[0]{\begin{document}}

\newcommand{\EndNoBibArticle}[0]{\end{document}}
\newcommand{\EndArticle}[0]{\bibliography{myrefs}\bibliographystyle{unsrtnat}\end{document}}

\newcommand{\pref}[1]{(\ref{#1})}

\label{chap:gramSchmidtLorentz}
\blogpage{http://sites.google.com/site/peeterjoot/math2011/gramSchmidtLorentz.pdf}
%\date{April 14, 2011}
\revisionInfo{gramSchmidtLorentz.tex}
%-----------------------------------------------------------------------------------

%\journalinfo{ApJ Letters, in press}
%\submitted{}
%\received{June 16, 2010} \accepted{October 20, 2010}
%\shorttitle{short title goes here.}

% these two need generalization.  have to switch order with journal format.
\beginArtNoToc
\chapter{Change of basis and Gram-Schmidt orthonormalization in special relativity}

\author{Peeter Joot \altaffilmark{1}}
\altaffiltext{1}{peeter.joot@utoronto.ca}

\begin{abstract}

%The standard language for the teaching and study of special relativity is that of tensor algebra.
While an explicit basis is common in the study of Euclidean spaces, it is usually implied in the study of inertial relativistic systems.  
There are some conceptual advantages to including the basis in the study of special relativistic systems.  
A Minkowski metric implies a non-orthonormal basis, and to deal with this complexity the 
concepts of reciprocal basis and the vector dual are introduced.  It is shown how the reciprocal basis is related to upper and lower index coordinate extraction, the metric tensor, change of basis, projections in non-orthonormal bases, and finally the Gram-Schmidt procedure.
%As with matrix representations of four vectors, the use of an explicit basis allows the details of the coordinate representation to be suppressed, allowing four vectors to be manipulated as complete entities in an inner products space.
It will be shown that Lorentz transformations can be viewed as change of basis operations.  The Lorentz boost in one spatial dimension will be derived using the Gram-Schmidt orthonormalization algorithm, and it will be shown how other Lorentz transformations can be derived using the Gram-Schmidt procedure.

\end{abstract}

\keywords{Lorentz boost, change of basis, special relativity, inertial frame, reciprocal basis, dual vector, Gram-Schmidt orthonormalization.}

\section{Abstract}

%\section{Introduction}

\section{Preliminaries, notation, and definitions}

\subsection{Four vectors, and the standard basis}

Four vectors will be written as a tuples of time and space coordinates.  These will be represented herein as non-bold letters of the form

\begin{equation}\label{eqn:gramSchmidtLorentz:10}
x = (c t, \Bx) = ( c t, x, y, z ) = ( x^0, x^1, x^2, x^3 ),
\end{equation}

where bold letters will be reserved for Euclidean vectors.  As usual, the factor of $c$ in the time coordinate is included so that the units of any of the coordinates in the tuple have dimensions of distance.  With

\begin{equation}\label{eqn:gramSchmidtLorentz:70}
\begin{aligned}
e_0 &= (1, 0, 0, 0) \\
e_1 &= (0, 1, 0, 0) \\
e_2 &= (0, 0, 1, 0) \\
e_3 &= (0, 0, 0, 1)
\end{aligned}
\end{equation}

the ordered set $\{e_0, e_1, e_2, e_3\}$ will be referred to as the standard basis.  Upper indexes will be used for the coordinates of the four vector in the standard basis (so $x^2$ is the 2-indexed coordinate of the four vector and not the square of $x$).  Lower indexed four vector coordinates will be introduced later once the reciprocal basis is introduced.

Repeated mixed upper and lower indexes will imply summation, with Greek indexes used for temporal and spatial indexes $\{0, 1, 2, 3\}$, and Latin indexes used in a Euclidean context $\{1, 2, \cdots N\}$.

\subsection{Relativistic inner product}

At the heart of special relativity is the definition of the invariant length or interval, which defines a distance like measure between a pair of vectors, relating both time and space coordinates.  This invariance may be codified by defining an inner product for the spacetime vector space of the form

\begin{equation}\label{eqn:gramSchmidtLorentz:50}
x \cdot y = \pm( x^0 y^0 - \Bx \cdot \By ).
\end{equation}

Here $y = (y^0, \By)$.  This is the Minkowski inner product, and is non-positive definite.  Opposite signs required for the spatial and temporal portions of the product, but the overall sign is arbitrary and conventions vary by author.  A positive sign will be used herein.
No attempt to motivate why a mixed sign for the time and space coordinates will be made here.  That more difficult job is deferred to any number of books covering special relativity
(e.g. \citep{landau1980classical}.)
%\citep[e.g.][]{landau1980classical}.

The use of a non-orthonormal basis, even in Euclidean spaces, makes life a bit more difficult.  There is, however, no choice in the matter for special relativity, since the standard basis is pseudo-orthonormal with elements \pref{eqn:gramSchmidtLorentz:70} unity only up to a sign.  For example, with the overall sign of the inner product \pref{eqn:gramSchmidtLorentz:50} chosen to be positive

\begin{equation}\label{eqn:gramSchmidtLorentz:860}
e_1 \cdot e_1 = e_2 \cdot e_2 = e_3 \cdot e_3 = -( e_0 \cdot e_0 ) = -1.
\end{equation}

A relativistic basis cannot be constructed for which all the basis vectors have strictly unit norm.  Unit vector will be used here loosely to refer to any vector $u$ such that $u \cdot u = \pm 1$.

\subsection{Reciprocal basis, duality, and coordinate representation with a non-orthonormal basis}

It is convenient to introduce the concept of a reciprocal basis when dealing with non-orthonormal spaces.  The utility of a reciprocal basis is not limited to the non-Euclidean vector space of special relativity.  The reciprocal basis elements are defined implicitly such that

\begin{equation}\label{eqn:gramSchmidtLorentz:130}
e_\alpha \cdot e^\beta = {\delta_\alpha}^\beta.
\end{equation}

The vector $e^\alpha$ is referred to as the dual of $e_\alpha$, and the ordered set of vectors $\{e^\alpha\}$ is called the reciprocal basis $\{e_\alpha\}$.

Given a coordinate representation

\begin{equation}\label{eqn:gramSchmidtLorentz:880}
x = x^\alpha e_\alpha,
\end{equation}

the coordinates may be extracted by taking dot products with the reciprocal basis elements

\begin{equation}\label{eqn:gramSchmidtLorentz:340}
x \cdot e^\alpha = (x^\beta e_\beta) \cdot e^\alpha = x^\beta {\delta_\beta}^\alpha = x^\alpha.
\end{equation}

Similarly, for the same vector $x$ represented in the reciprocal basis

\begin{equation}\label{eqn:gramSchmidtLorentz:900}
x = x_\alpha e^\alpha,
\end{equation}

the coordinates may be extracted by taking dot products with $e_\alpha$

\begin{equation}\label{eqn:gramSchmidtLorentz:340b}
x \cdot e_\alpha = (x_\beta e^\beta) \cdot e_\alpha = x_\beta {\delta^\beta}_\alpha = x_\alpha.
\end{equation}

In this context there is nothing special about either upper or lower indexes.  They are just coordinates with respect to a basis and its reciprocal basis, respectively.  When the original basis happens to be orthonormal, there is equality between the basis vectors and their reciprocal duals $e_\alpha = e^\alpha$, as well as between the coordinates calculated from those bases respectively $x_\alpha = x^\alpha$.  In tensor algebra, upper indexes are ``special'' since the invariant transformations are defined in terms of those coordinates, but that is really just a choice of basis.  It is in fact possible \citep{doran2003gap} to express Lorentz transformations in a completely coordinate free fashion, where there is freedom to employ upper or lower index representation of the coordinates, or coordinates with respect to any basis, even one that is not normal.

\subsubsection{Reciprocal basis example in 2D Euclidean space}

The calculation of the reciprocal basis elements may be dependent on the complete set of elements in the non-dual basis.  This can be illustrated nicely by considering an example of an oblique basis in a Euclidean space.

For a 2 dimensional Euclidean space, a non-orthonormal basis $A = \{e_1, e_2\}$ such as

\begin{equation}\label{eqn:gramSchmidtLorentz:90}
e_1 =
\begin{bmatrix}
1 \\
1
\end{bmatrix},\qquad
e_2 =
\begin{bmatrix}
1 \\
2
\end{bmatrix},
\end{equation}

may be chosen.  In this column vector representation the duality relation \pref{eqn:gramSchmidtLorentz:130} takes the form

\begin{equation}\label{eqn:gramSchmidtLorentz:150}
\begin{bmatrix}
{e_1}^\T \\
{e_2}^T
\end{bmatrix}
\begin{bmatrix}
e^1 & e^2
\end{bmatrix} = I.
\end{equation}

Inversion provides the dual vectors
\begin{equation}\label{eqn:gramSchmidtLorentz:170}
\begin{bmatrix}
e^1 & e^2
\end{bmatrix}
=
{
\begin{bmatrix}
{e_1}^\T \\
{e_2}^T
\end{bmatrix}
}^{-1}
=
{\begin{bmatrix}
1 & 1 \\
1 & 2
\end{bmatrix}}^{-1}
=
\begin{bmatrix}
2 & -1 \\
-1 & 1
\end{bmatrix},
\end{equation}

or

\begin{equation}\label{eqn:gramSchmidtLorentz:190}
e^1 =
\begin{bmatrix}
2 \\
-1
\end{bmatrix}, \qquad
e^2 =
\begin{bmatrix}
-1 \\
1
\end{bmatrix}.
\end{equation}

The problem of solving for the coordinates $a,b$ of a vector $x = a e_1 + b e_2$ in this oblique basis now reduces to taking dot products

\begin{subequations}\label{eqn:gramSchmidtLorentz:920}
\begin{align}
x \cdot e^1 &= a e_1 \cdot e^1 + b \cancel{e_2 \cdot e^1} = a \\
x \cdot e^2 &= a \cancel{e_1 \cdot e^2} + b e_2 \cdot e^2 = b.
\end{align}
\end{subequations}

As a concrete example consider

\begin{equation}\label{eqn:gramSchmidtLorentz:230}
\begin{aligned}
x &=
\begin{bmatrix}
4 \\
2
\end{bmatrix}  \\
&=
\left(
\begin{bmatrix}
4 \\
2
\end{bmatrix}
\cdot e^1
\right)
e_1
+
\left(
\begin{bmatrix}
4 \\
2
\end{bmatrix}
\cdot e^2
\right)
e_2 \\
&= 6
\begin{bmatrix}
1 \\
1
\end{bmatrix}
+
-2
\begin{bmatrix}
1 \\
2
\end{bmatrix}.
\end{aligned}
\end{equation}

Coordinates may also be computed with respect to the reciprocal basis.  With $x = c e^1 + d e^2$, dotting with $e_1$ and $e_2$ respectively provides these

\begin{equation}\label{eqn:gramSchmidtLorentz:360}
\begin{aligned}
x \cdot e_1 &= c e^1 \cdot e_1 + d \cancel{e^2 \cdot e_1} = c \\
x \cdot e_2 &= c \cancel{e^1 \cdot e_2} + d e^2 \cdot e_2 = d.
\end{aligned}
\end{equation}

Again considering the concrete example above

\begin{equation}\label{eqn:gramSchmidtLorentz:290}
\begin{aligned}
x
&=
\begin{bmatrix}
4 \\
2
\end{bmatrix} \\
&=
\left(
\begin{bmatrix}
4 \\
2
\end{bmatrix}
\cdot e_1
\right)
e^1
+
\left(
\begin{bmatrix}
4 \\
2
\end{bmatrix}
\cdot e_2
\right)
e^2 \\
&= 6
\begin{bmatrix}
2 \\
-1
\end{bmatrix}
+
8
\begin{bmatrix}
-1 \\
1
\end{bmatrix}.
\end{aligned}
\end{equation}

\begin{figure}[htp]
\centering
\includegraphics[totalheight=0.3\textheight]{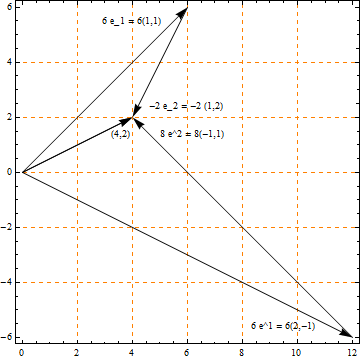}
\caption{Vector projections in oblique and reciprocal frames.}\label{fig:obliqueReciprocal}
\end{figure}

This pair of coordinate calculations is depicted in figure (\pref{fig:obliqueReciprocal}).

The projections onto the elements of a general non-orthonormal basis are

\begin{subequations}\label{eqn:gramSchmidtLorentz:940}
\begin{align}
\Proj_{e_i}(x) &= (x \cdot e^i) e_i \\
\Proj_{e^i}(x) &= (x \cdot e_i) e^i,
\end{align}
\end{subequations}

(no sum) and not 

\begin{equation}\label{eqn:gramSchmidtLorentz:960}
\Proj_{e_i}(x) = \Proj_{e^i}(x) = \frac{x \cdot e_i}{e_i \cdot e_i} e_i.
\end{equation}

The operator \pref{eqn:gramSchmidtLorentz:940} for projecting onto elements of non-orthonormal bases does imply \pref{eqn:gramSchmidtLorentz:960} for the special cases of orthonormal and pseudo-orthonormal bases, since $e^i = e_i/(e_i \cdot e_i)$.

\subsubsection{Projections}

Examples of projections onto the Euclidean non-orthonormal basis above have been seen.  In general the relations \pref{eqn:gramSchmidtLorentz:340}, and \pref{eqn:gramSchmidtLorentz:340b} allow for such Fourier decomposition of a vector into components in each of the respective basis directions

\begin{equation}\label{eqn:gramSchmidtLorentz:330}
\begin{aligned}
x &= x^\alpha e_\alpha = (x \cdot e^\alpha) e_\alpha \\
  &= x_\alpha e^\alpha = (x \cdot e_\alpha) e^\alpha.
\end{aligned}
\end{equation}

With \pref{eqn:gramSchmidtLorentz:330} containing $x$ on both the LHS and in the RHS as $(x \cdot e^\alpha) e_\alpha$, this relation has an appearance of being somewhat recursive.  This is however, an important property, since each of the RHS terms represents a projection.  The projection of a vector onto the basis element $e_\alpha$ is

\begin{equation}\label{eqn:gramSchmidtLorentz:560}
\Proj_{e_\alpha}(x) = (x \cdot e^\alpha) e_\alpha,
\end{equation}

(no sum implied.)  
This will be important since the Gram-Schmidt procedure is essentially just the repeated subtraction of projections, and knowledge of how to express projections for a non-orthonormal basis is required.

\subsubsection{Gram-Schmidt procedure generalized to non-orthonormal bases}

Aside for some additional care required to express projections, the Gram-Schmidt procedure is the same as in Euclidean space.
Given a set of mutually normal unit vectors $\{f_0, \cdots, f_\alpha\}$, the set may be extended by an additional normal vector.  Provided a vector $a$ lying outside of the span of this set can be found, subtraction of the projections of $a$ from the all the elements of this set leaves only the component of $a$ normal to all vectors in this set.  That is

\begin{equation}\label{eqn:gramSchmidtLorentz:660}
\begin{aligned}
b
&= a - \sum_{\sigma \le \alpha} \Proj_{f_\sigma}(a) \\
&= a - \sum_{\sigma \le \alpha} (a \cdot f_\sigma) f^\sigma.
\end{aligned}
\end{equation}

Normalization $f_{\alpha+1} = b/\sqrt{\Abs{b \cdot b}}$ allows the set to be extended by an additional unit vector.  This process can be repeated until a complete basis is formed.

\subsubsection{Reciprocal basis for relativity}

Because it is not possible to have a strictly orthonormal basis in a relativistic context, the reciprocal basis must have a place in the geometry of relativity.
It is easily verified that the vectors

\begin{equation}\label{eqn:gramSchmidtLorentz:70b}
\begin{aligned}
e^0 &= (1, 0, 0, 0) \\
e^1 &= (0, -1, 0, 0) \\
e^2 &= (0, 0, -1, 0) \\
e^3 &= (0, 0, 0, -1)
\end{aligned}
\end{equation}

are dual to the standard basis elements \pref{eqn:gramSchmidtLorentz:70} according to the definition \pref{eqn:gramSchmidtLorentz:130}.

\subsubsection{Metric tensors}

Upper and lower index coordinates with respect to any basis and its reciprocal, orthonormal or not, are related by dot products of the basis elements, and are not independent.  Given a vector with both upper and lower index representation

\begin{equation}\label{eqn:gramSchmidtLorentz:980}
x = x^\alpha e_\alpha = x_\beta e^\beta,
\end{equation}

utilizing the coordinate representation in the chosen basis, and in the reciprocal basis, the dot product in terms of coordinates is found to take the standard tensor form

\begin{equation}\label{eqn:gramSchmidtLorentz:1060}
x \cdot x = (x^\alpha e_\alpha) \cdot (x_\beta e^\beta) = x^\alpha x_\beta {\delta^\alpha}_\beta = x^\alpha x_\alpha.
\end{equation}

The upper and lower coordinates may be related by taking dot products with $e_\mu$, and $e^\mu$ as follows

\begin{subequations}\label{eqn:gramSchmidtLorentz:1000}
\begin{align}
x_\mu &= x \cdot e_\mu = (e_\mu \cdot e_\alpha) x^\alpha \\
x^\mu &= x \cdot e^\mu = (e^\mu \cdot e^\beta) x_\beta.
\end{align}
\end{subequations}

These pairs of dot products define the metric tensors for the pair of bases

\begin{subequations}
\label{eqn:gramSchmidtLorentz:1020}
\begin{align}
g_{\mu \nu} &= e_\mu \cdot e_\nu \\
g^{\mu \nu} &= e^\mu \cdot e^\nu,
\end{align}
\end{subequations}

which provide the raising and lowering operations in their tensor form

\begin{subequations}
\label{eqn:gramSchmidtLorentz:1040}
\begin{align}
x_\mu &= g_{\mu \nu} x^\nu \\
x^\mu &= g^{\mu \nu} x_\nu.
\end{align}
\end{subequations}

From \pref{eqn:gramSchmidtLorentz:1040} observe that the dot product \pref{eqn:gramSchmidtLorentz:1060} can be written in terms of the metric tensor

\begin{equation}\label{eqn:gramSchmidtLorentz:1041}
x \cdot x = g_{\mu \nu} x^\mu x^\nu.
\end{equation}

Like any other vector, a basis vector can be split into its Fourier components

\begin{subequations}
\label{eqn:gramSchmidtLorentz:1161}
\begin{align}
e^\alpha &= (e^\alpha \cdot e^\beta) e_{\beta} \\
e_\alpha &= (e_\alpha \cdot e_\beta) e^{\beta}.
\end{align}
\end{subequations}

These are, respectively, $e^\alpha = g^{\alpha\beta} e_\beta$ and $e_\alpha = g_{\alpha \beta} e^\beta$, demonstrating that the metric tensor can be used to raise or lower the basis vectors just like coordinates.

The metric tensors \pref{eqn:gramSchmidtLorentz:1020} are generally basis dependent and not diagonal or identical.  For orthonormal and pseudo-orthonormal bases $g^{\mu\nu} = g_{\mu\nu}$ and are also diagonal.
The dot product itself is not basis dependent, and will produce the same result for any basis and its associated coordinates.  The Lorentz separation, defined in terms of the fundamental mixed sign relationship \pref{eqn:gramSchmidtLorentz:50}, will be identical for all bases, even ones where the basis vectors are not normal.

\subsubsection{Change of basis}

Consider a vector with coordinate representations in a pair of bases, not necessarily orthonormal

\begin{equation}\label{eqn:gramSchmidtLorentz:1080}
x = y^\alpha f_\alpha = x^\beta e_\beta.
\end{equation}

Taking dot products with reciprocal frame elements relates the coordinates

\begin{subequations}
\label{eqn:gramSchmidtLorentz:1100}
\begin{align}
y^\mu &= (f^\mu \cdot e_\nu ) x^\nu \\
x^\mu &= (e^\mu \cdot f_\nu) y^\nu.
\end{align}
\end{subequations}

Defining tensors for the various dot product combinations

\begin{subequations}
\label{eqn:gramSchmidtLorentz:1101}
\begin{align}
{\wedge^\mu}_\nu &= f^\mu \cdot e_\nu \\
\wedge^{\mu\nu} &= f^\mu \cdot e^\nu \\
{\wedge_\mu}^\nu &= f_\mu \cdot e^\nu \\
\wedge_{\mu \nu} &= f_\mu \cdot e_\nu 
\end{align}
\end{subequations}

allows for the coordinate transformations of \pref{eqn:gramSchmidtLorentz:1100} to take their more conventional tensor form

\begin{subequations}
\label{eqn:gramSchmidtLorentz:1140}
\begin{align}
y^\mu &= {\wedge^\mu}_\nu x^\nu \\
x^\mu &= 
%g^{\alpha \mu} g_{\beta \nu} {\wedge^\beta}_\alpha
{\wedge_\nu}^\mu
y^\nu.
\end{align}
\end{subequations}

Observe that the invariant length as seen in standard tensor form is necessarily preserved by a change of basis transformation

\begin{subequations}
\label{eqn:gramSchmidtLorentz:1160}
\begin{align}
y^\mu y_\mu &= (y^\mu f_\mu) \cdot (y_\nu f^\nu) = x \cdot x \\
x^\mu x_\mu &= (x^\mu e_\mu) \cdot (x_\nu e^\nu) = x \cdot x.
\end{align}
\end{subequations}

It is thus natural to consider the coordinates $y^\alpha$ after transformation as the same vector that had the coordinates $x^\alpha$.  They are just representations under different bases.

\subsubsection{Determination of the transformed basis from the transformation matrix.}

Forming projections with respect to the standard basis provides the coordinates of the transformed frame

\begin{equation}\label{eqn:gramSchmidtLorentz:1200}
f^\mu = (f^\mu \cdot e^\nu) e_\nu = \wedge^{\mu\nu} e_\nu.
\end{equation}

This can also be viewed as a contraction of the transformation matrix \pref{eqn:gramSchmidtLorentz:1101} with $e^\beta$

\begin{equation}\label{eqn:gramSchmidtLorentz:1201}
{\wedge^\mu}_\nu e^\nu = (f^\mu \cdot e_\nu) e^\nu = f^\mu.
\end{equation}

Written out in full, the transformed coordinates are

\begin{subequations}
\label{eqn:gramSchmidtLorentz:1220}
\begin{align}
f_\mu &= ( \wedge_{\mu 0}, \wedge_{\mu 1}, \wedge_{\mu 2}, \wedge_{\mu 3} ) \\
f^\mu &= ( \wedge^{\mu 0}, \wedge^{\mu 1}, \wedge^{\mu 2}, \wedge^{\mu 3} ).
\end{align}
\end{subequations}

%\subsubsection{Basis orientation.}
%
%As in Euclidean space, a determinant of the basis vector coordinates provides an orientation for the basis.  The standard basis, for example, regardless of the metric can be defined as a reference orientation and has a unit determinant.  With a row vector matrix representation of the coordinates, that is
%
%\begin{equation}\label{eqn:gramSchmidtLorentz:680}
%\begin{vmatrix}
%[e_0] \\
%[e_1] \\
%[e_2] \\
%[e_3]
%\end{vmatrix}
%=
%\begin{vmatrix}
%1 & 0 & 0 & 0 \\
%0 & 1 & 0 & 0 \\
%0 & 0 & 1 & 0 \\
%0 & 0 & 0 & 1 \\
%\end{vmatrix}
%= 1.
%\end{equation}
%
%For the basis associated with a Lorentz transformation, the sign of the basis elements or the order within the basis can be adjusted so that the determinants of the coordinates of the unit vectors is unity.  This will ensure a positive orientation with respect to the standard basis, and avoids non-physical transformations involving reflection in time or space.

\section{Relativity}
\subsection{Proper separation}

Given a particle parametrization along a trajectory $x(\lambda)$ in spacetime, the average spacetime length between a pair of points on this path can be computed

\begin{equation}\label{eqn:gramSchmidtLorentz:400}
s_b - s_a = \int_{\lambda = a}^b \sqrt{ \frac{d x(\lambda)}{d\lambda} \cdot \frac{d x(\lambda)}{d\lambda} } d\lambda.
\end{equation}

A trajectory $x(\lambda)$ may be reparametrized in terms of the instantaneous proper separation $s$.  The first derivative of $x(s)$ with respect to $s$ is a timelike unit vector ($x' \cdot x' = e_0 \cdot e_0 = 1$) along any point of the curve.

This is nicely demonstrated by example.

Consider an inertial system, with a particle moving along a constant velocity trajectory, parametrized by an external observers time $t$

\begin{equation}\label{eqn:gramSchmidtLorentz:440}
x(t) = (c t, \Bv t)
\end{equation}

The proper separation anywhere along this spacetime curve is

\begin{equation}\label{eqn:gramSchmidtLorentz:460}
\begin{aligned}
s
&= \int \sqrt{(c t, \Bv t) \cdot (c t, \Bv t)} dt  \\
&= \int \sqrt{ c^2 - \Bv^2 } dt  \\
&= \sqrt{ c^2 - \Bv^2 } t
\end{aligned}
\end{equation}

Proper length reparametrization of this path is thus

\begin{equation}\label{eqn:gramSchmidtLorentz:480}
x(s) = \inv{\sqrt{c^2 - \Bv^2}} (c, \Bv) s,
\end{equation}

The derivative

\begin{equation}\label{eqn:gramSchmidtLorentz:500}
\frac{dx}{ds} = \inv{\sqrt{c^2 - \Bv^2}} (c, \Bv),
\end{equation}

is now easily observed to be of unit length

\begin{equation}\label{eqn:gramSchmidtLorentz:520}
\frac{dx}{ds} \cdot \frac{dx}{ds} 
= \inv{c^2 - \Bv^2} (c^2 - \Bv^2) = 1.
\end{equation}

For an inertial system, where $d^2 x/ds^2 = 0$ there is only a timelike component for the trajectory when parametrized by proper length.  That is

\begin{equation}\label{eqn:gramSchmidtLorentz:580}
x(s) = \frac{dx}{ds} s.
\end{equation}

In general a proper length differential of a trajectory is parametrized by the elapsed time in the frame for which the particle is instantaneously at rest.  A student of special relativity is familiar with being able to switch to a frame in which the particle is instantaneously at rest by performing a Lorentz boost.  This gives a first indirect hint that such a transformation can be interpreted as nothing more than a change of basis.

\subsection{Lorentz boost as a change of basis}

For the trajectory \pref{eqn:gramSchmidtLorentz:440} it was found that the proper length derivative was a timelike unit vector for the frame in which the particle was at rest.  That unit vector can be used as part of a basis for that rest frame.  Once this basis is completed with its spatial unit vectors, it will be seen how the particle's rest basis and an observer basis are related by Lorentz transformation.

\subsubsection{Illustration by example.  One spatial dimension}

For simplicity, consider a two dimensional spacetime vector space, with a particle trajectory in an inertial frame parametrized by its proper length

\begin{equation}\label{eqn:gramSchmidtLorentz:600}
x(s) = \gamma (1, \beta) s.
\end{equation}

Labeling this time like unit vector in the particle's rest frame $f_0$ provides the first element of a basis in the particle's rest frame

\begin{equation}\label{eqn:gramSchmidtLorentz:620}
f_0 = \frac{dx}{ds} = \gamma (1, \beta) = f^0.
\end{equation}

The particle's trajectory in the rest frame, in terms of the basis to be determined is thus

\begin{equation}\label{eqn:gramSchmidtLorentz:640}
x(s) = s f_0 + 0 f_1.
\end{equation}

The task is to compute this basis $\{f_0, f_1\}$ for the particle's rest frame.

For this one dimensional spatial example, any vector lying outside of the span of $\{f_0\}$ can be picked to find an additional vector normal to that.  One such vector, chosen arbitrarily, is $e_1 = (0, 1)$,

\begin{align*}
b
&= e_1 - (e_1 \cdot f_0) f^0 \\
&= (0, 1) - (0, 1) \cdot (1, \beta) \gamma^2 (1, \beta) \\
&= (0, 1) + \beta \gamma^2 (1, \beta) \\
&= (\gamma^2 \beta, 1 + \beta^2 \gamma^2) \\
&= \gamma^2 (\beta, 1 ).
\end{align*}

This can be normalized as either $\pm \gamma(\beta, 1)$.  The positive choice ensures that the determinant of the coordinates matches that of the standard basis (i.e. unity).

\begin{equation}\label{eqn:gramSchmidtLorentz:685}
\begin{vmatrix}
[f_0] \\
[f_1]
\end{vmatrix}
=
\begin{vmatrix}
\gamma & \gamma \beta \\
\gamma \beta & \gamma
\end{vmatrix}
=
\gamma^2 ( 1 - \beta^2)
= 1.
\end{equation}

This unit determinant ensures that the particle's rest frame basis has the same orientation as the standard basis, involving no reflections in space or time.

Using \pref{eqn:gramSchmidtLorentz:1101}, the change of basis matrix from the standard basis of a stationary observer, to the rest frame basis for the particle, is

\begin{equation}\label{eqn:gramSchmidtLorentz:720}
\begin{aligned}
\Norm{ {\wedge^\mu}_\nu }
%&= f^\mu \cdot e_\nu \\
&=
\begin{bmatrix}
f^0 \cdot e_0
&
f^0 \cdot e_1
\\
f^1 \cdot e_0
 &
f^1 \cdot e_1
\end{bmatrix} \\
&=
\begin{bmatrix}
( 1, \beta )
\cdot
(1, 0)
\gamma &
( 1, \beta )
\cdot
(0, 1)
\gamma \\
( -\beta, -1)
\cdot
(1, 0)
\gamma &
( -\beta, -1)
\cdot
(0, 1)
\gamma
\end{bmatrix} \\
&=
\gamma
\begin{bmatrix}
1 & - \beta \\
- \beta & 1
\end{bmatrix}.
\end{aligned}
\end{equation}

The Lorentz boost matrix for a one dimensional motion along the $\gamma c t (1, \beta)$ trajectory has been derived by considering only geometry.

Considering the Lorentz boost matrix above provides a physical justification for the demand that the determinant of the transformed coordinates is unity.  Had we chosen a basis $\{f_0, -f_1\}$ with $f_0$ and $f_1$ as defined above, our coordinates would transform as

\begin{equation}\label{eqn:gramSchmidtLorentz:721}
\begin{bmatrix}
ct' \\
x'
\end{bmatrix}
=
\gamma
\begin{bmatrix}
1 & \beta \\
-1 & -\beta
\end{bmatrix}
\begin{bmatrix}
ct \\
x
\end{bmatrix}.
\end{equation}

These coordinates still have the $(ct')^2 - (x')^2 = (ct)^2 - x^2$ Lorentz invariance, but in the $\beta \rightarrow 0$ case produce a spatial reflection

\begin{equation}\label{eqn:gramSchmidtLorentz:722}
\begin{bmatrix}
ct' \\
x'
\end{bmatrix}
=
\begin{bmatrix}
ct \\
-x
\end{bmatrix}.
\end{equation}

\begin{figure}[htp]
\centering
\includegraphics[totalheight=0.3\textheight]{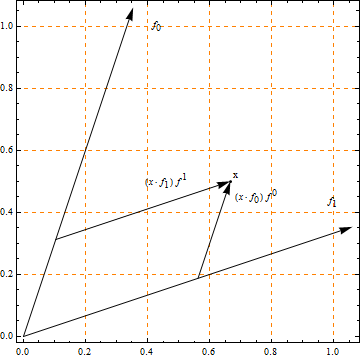}
\caption{Minkowski diagram utilizing reciprocal frame projections.}\label{fig:minkowskiReciprocal}
\end{figure}

The geometry of the reciprocal frame makes the calculation of Minkowski diagrams simple.  Figure \pref{fig:minkowskiReciprocal} illustrates a plot with $\beta = 1/3$ of boosted basis vectors $f_0, f_1$ and the projections $(x \cdot f_0) f^0, (x \cdot f_1) f^1$ of a vector onto the rest frame basis for the particle at rest.

\subsubsection{On uniqueness}

A two dimensional boost of speed $c \beta$ along the spatial direction $(\cos\theta, \sin\theta)$ can be shown to have the transformation matrix

\begin{equation}\label{eqn:gramSchmidtLorentz:1300}
%\Norm{ {\wedge^\mu}_\nu }
%=
\begin{bmatrix}
\gamma & -\gamma \beta \cos\theta & -\gamma \beta \sin\theta \\
-\gamma \beta \cos\theta & 1 + (\gamma -1) \cos^\theta & (\gamma -1) \cos\theta \sin\theta \\
-\gamma \beta \sin\theta & (\gamma -1) \sin\theta \cos\theta & 1 + (\gamma -1) \sin\theta
\end{bmatrix}.
\end{equation}

Utilizing \pref{eqn:gramSchmidtLorentz:1220} the particle's rest frame basis is found to be

\begin{subequations}
\label{eqn:gramSchmidtLorentz:1320}
\begin{align}
f_0 &= \gamma ( 1, \beta \cos\theta, \beta\sin\theta ) \\
f_1 &= (\gamma \beta \cos\theta, 1 + (\gamma -1) \cos^\theta, (\gamma -1) \cos\theta \sin\theta) \\
f_2 &= (\gamma \beta \sin\theta, (\gamma -1) \sin\theta \cos\theta, 1 + (\gamma -1) \sin\theta).
\end{align}
\end{subequations}

This is the Lorentz transform matrix corresponding to a particle trajectory of

\begin{equation}\label{eqn:gramSchmidtLorentz:800}
x(s) = \gamma s ( 1, \beta \cos\theta, \beta \sin\theta),
\end{equation}

for which the timelike unit vector is $f_0 = ( 1, \beta \cos\theta, \beta \sin\theta )$.  With more than one spatial direction, the boost matrix, or the corresponding basis in the particle's rest frame, cannot be uniquely determined using the Gram-Schmidt procedure used above in one spatial dimension.  For example, seeding the Gram-Schmidt procedure starting with this timelike unit vector and picking $e_0 = (1, 0, 0)$ as the first vector outside of the span of $\{f_0\}$, the following positively oriented unit normalized basis can be calculated

\begin{subequations}
\label{eqn:gramSchmidtLorentz:820}
\begin{align}
f_0 &= \gamma ( 1, \beta \cos\theta, \beta\sin\theta ) \\
f_1 &= \gamma ( \beta, \cos\theta, \sin\theta ) \\
f_2 &= \gamma ( 0, -\sin\theta, \cos\theta ).
\end{align}
\end{subequations}

The matrix of this linear transformation is

\begin{equation}\label{eqn:gramSchmidtLorentz:840}
\Norm{ {\wedge^\mu}_\nu } =
\begin{bmatrix}
\gamma & - \gamma \beta \cos\theta & - \gamma \beta \sin\theta \\
-\gamma \beta & \gamma \cos\theta & \gamma \sin\theta \\
0 & -\sin\theta & \cos\theta
\end{bmatrix}.
\end{equation}

While this has unit determinant, and necessarily preserves the invariant length of a vector, it does not have the symmetric form of the boost associated with the spatial velocity $c \beta (\cos\theta, \sin\theta)$.

It is, of course, possible to determine the basis associated with any Lorentz transformation matrix using equations \pref{eqn:gramSchmidtLorentz:1220}.  For example, for the two spatial direction boost matrix

\begin{equation}\label{eqn:gramSchmidtLorentz:1400}
\begin{bmatrix}
\gamma & - \gamma \beta \cos\theta & -\gamma \beta \sin\theta \\
-\gamma \beta \cos\theta & 1 + ( \gamma -1) \cos^2\theta & (\gamma -1) \sin\theta \cos\theta \\
-\gamma \beta \sin\theta & (\gamma -1) \sin\theta \cos\theta & 1 + (\gamma -1) \sin^2\theta \\
\end{bmatrix}.
\end{equation}

From these the basis vectors in the particle's rest frame follow

\begin{subequations}
\begin{align}\label{eqn:gramSchmidtLorentz:1420}
f_0 &= 
(\gamma , \gamma \beta \cos\theta , \gamma \beta \sin\theta) \\
f_1 &=
(\gamma \beta \cos\theta , 1 + ( \gamma -1) \cos^2\theta , (\gamma -1) \sin\theta \cos\theta) \\
f_2 &=
(\gamma \beta \sin\theta , (\gamma -1) \sin\theta \cos\theta , 1 + (\gamma -1) \sin^2\theta).
\end{align}
\end{subequations}

%\section{Discussion}
%\acknowledgments

\section{Conclusion}

Results are summarized as follows
\begin{enumerate}
\item The concepts of reciprocal basis and vector dual have been defined.
\item Upper and lower index coordinates are defined as dot products with the chosen and reciprocal basis respectively.
\item The metric tensor has been defined in terms of dot products of the basis vectors.
\item Projection and the Gram-Schmidt procedure for a non-orthonormal basis has been detailed.
\item A worked example illustrating reciprocal basis and projection for a non-orthonormal (Euclidean) basis has been provided.
\item It is shown how a Lorentz transformation can be described as a change of basis.
\item How to calculate the basis and the reciprocal basis for an arbitrary Lorentz transformation has been detailed.
\item A derivation of a one dimensional boost using only the Gram-Schmidt procedure has been provided.
\end{enumerate}

Some of the concepts used herein, especially that of the reciprocal basis, have been borrowed from the context of Geometric Algebra, where coordinate free methods are developed in considerably more depth and generality.

Without the learning curve of attempting a study of Geometric Algebra, an attempt has been made to illustrate some of the conceptual advantages of including the basis in the study of special relativity, even for inertial frames where the basis is usually omitted.  The intent has been to detail the mathematical tools required in a structured and standalone fashion so that the student can then proceed to apply additional ideas and tools of Euclidean vector algebra to the study of special relativity.

\EndArticle